# Hot exciton relaxation in coupled ultra-thin CdTe/ZnTe quantum well structures

V. Agekyan[a], G. Budkin[c], M. Chukeev[a,c], N. Filosofov[a], G. Karczewski[b], A. Serov[a] and A. Reznitsky[c,]*

[a] *Saint-Petersburg State University, St. Petersburg, 194034 Russia*
[b] *Institute of Physics PAN, Warsaw, PL-02-668 Poland*
[c] *Ioffe Institute, St. Petersburg, 194021 Russia*

**Abstract:** The photoluminescence (PL) and PL excitation (PLE) spectra of CdTe/ZnTe asymmetric double quantum well (QW) structures are studied on a series of samples containing two CdTe layers with nominal thicknesses of 2 and 4 monolayers (ML) in the ZnTe matrix. The samples differ in the thickness of the ZnTe spacer between CdTe QWs which is 45, 65 and 75 ML. It has been found that at above-barrier excitation the PL from a shallow QW at sufficiently weak excitation intensities is determined by recombination of hot excitons. It is shown that under these conditions, when PL is excited by lasers with different wavelengths, the ratio of the PL intensities from shallow and deep QWs decreases exponentially with an increase of the initial kinetic energy of hot excitons. It is found that energy relaxation of hot excitons with LO phonon emission determine the shape of the PLE spectrum of shallow QW in the range of exciton kinetic energies up to more than 20 LO phonons above ZnTe bandgap. We have shown that the results obtained are well described by the model of charge and energy transfer between QWs.

* Electronic address: alexander.reznitsky@mail.ioffe.ru

## 1. Introduction

The optical properties of CdTe/ZnTe heterostructures with ultrathin layers are the subject of numerous studies. However, for this system there is still no accurate data on the values of valence and conductive bands offsets [1–6]. Because the cation is replaced when crossing the interface, the barrier for hole should be significantly lower as compared to the electron. Another important circumstance is the significant mismatch between the lattice constants of ZnTe and CdTe, which strongly affects the real structure of CdTe quantum well (QW). Relaxation of mechanical stresses arising for this reason leads to the fact that the CdTe layer, starting from a certain nominal thickness, is a planar array of quantum dots (QDs) [7–9]. Most practical applications of such systems are based on the use of multilayer structures: the use of several layers containing self-organized quantum dots as an active area of optoelectronic devices significantly improves their characteristics if the electronic states in the layers are energetically coupled. Therefore, the study of tunneling processes in multilayer low-dimensional systems is interesting and important both from the point of view of their practical application and also from the point of view of understanding the fundamental properties of such systems.

One of the simplest objects for analyzing and characterizing the tunneling processes is asymmetric double QWs. The tunnel transparency of the barrier separating electronic states of QWs depends both on the nominal thicknesses of QWs and on the thickness of the barrier layer [10–13]. Moreover, the permeability of the barrier, as shown in [14], may be different for single electrons or holes and for excitons. It realizes the conditions under which a cascade relaxation of hot excitons with LO phonon emission is the main channel of energy transfer to the shallow QW in a wide range of photoexcitation energies.

A characteristic manifestation of such properties of coupled QWs is the dependence of the relative intensity of light emitted from shallow ($I_1$) and deep ($I_2$) QWs on the excitation intensity.

At above-barrier excitation, absorption of photons results in the formation in the barrier layer of separated electron-hole pairs and/or excitons with kinetic energy equals to ($E_0 - E_g$), where $E_0$ is the energy of excitation photons, and $E_g$ is the forbidden gap of the barrier material (hot carriers and excitons [15]). In II–VI compounds, photoexcited hot electrons and excitons cool down by emitting LO phonons, and are captured from the barrier layer to the levels of QWs followed by relaxation to a local potential minimum.

The capture of a single carrier, for example, an electron, into a shallow well leads to the realization of one of two possibilities: (i) the captured carrier waits the carrier of another sign (hole in our case) to be captured into the same potential well, after which the e–h pair can recombine; (ii) the electron has enough time to tunnel into a deep QW. Obviously, the realization of the latter possibility depends on the ratio of the tunneling time $t_t$ determined by the width of the potential barrier between the QWs, and the electron waiting time $t_w$ for a hole capture, which depends primarily on the excitation intensity.

The aim of this paper is to study the processes and mechanisms of charge and energy transfer in an asymmetric system of CdTe/ZnTe double QWs, to verify applicability of the existing theoretical model for describing its properties, and to test experimentally the relative role of hot excitons and free carries in the formation of the PL and PL excitation spectra of the systems under consideration.

## 2. Samples and experimental details

We have studied PL and PLE spectra, as well as the dependence on excitation intensity of the ratio of integral PL intensities of shallow and deep QWs $I_1/I_2$ in a series of samples containing two layers of CdTe $D_1$ and $D_2$ with nominal thickness of 2 and 4 monolayers (ML) in the ZnTe matrix. The samples were grown on a GaAs(100) substrate and included 1 μm thick ZnTe buffer layer, and two CdTe QWs separated by ZnTe barrier with thicknesses of 45, 65 and 75 ML. The entire structure was covered with a 100 nm thick ZnTe cap layer. The ZnTe layers were grown by the standard molecular beam epitaxy technics, while the CdTe QWs were synthesized in the atomic layer epitaxy mode. To excite the PL we used 442 nm He-Cd laser line, as well as radiation from a set of semiconductor lasers with wavelengths from 517 down to 405 nm. In all experiments, the maximum excitation power density did not exceed 10 W/cm$^2$, which means the creation of less than one e–h pair per CdTe QD. To study the PLE spectra, Xe lamp radiation, transmitted through an additional monochromator was used. The excitation power density under these conditions did not exceed 10 mW/cm$^2$.

## 3. Results and discussion

Low temperature PL spectra for above-barrier excitation include two bands $I_1$ and $I_2$ which are obviously caused by the recombination of excitons in the $D_1$ and $D_2$ layers. FWHM of PL bands from shallow and deep QWs is 3–4 meV and 15–20 meV, respectively. These values may mean that the $D_1$ layer is formed by more uniform ZnCdTe solid solution, while the $D_2$ layer appears to be the ensemble of self-organized QDs. One can see that the width of PL bands, is determined by the size dispersion of QDs, which increases with the amount of deposited CdTe [7–9].

An important feature of these PL spectra is the different dependence of the intensity of the $I_1$ and $I_2$ emission bands on the excitation intensity. As an example, Fig. 1 shows normalized low-temperature PL spectra of a sample with a barrier width of 75 ML obtained at two different intensities of above-barrier excitation. It worth to note that under such excitation PL spectra in the spectral range 2.36 eV < $E$ < 2.38 eV include also a few weak narrow lines, apparently, due to the recombination of excitons bound to shallow impurities in the ZnTe layers. The intensities of these lines depend on the level of excitation and are at least 100 times weaker than $I_1$. Further in the text we will not discuss their properties and will focus only on $I_1$ and $I_2$.

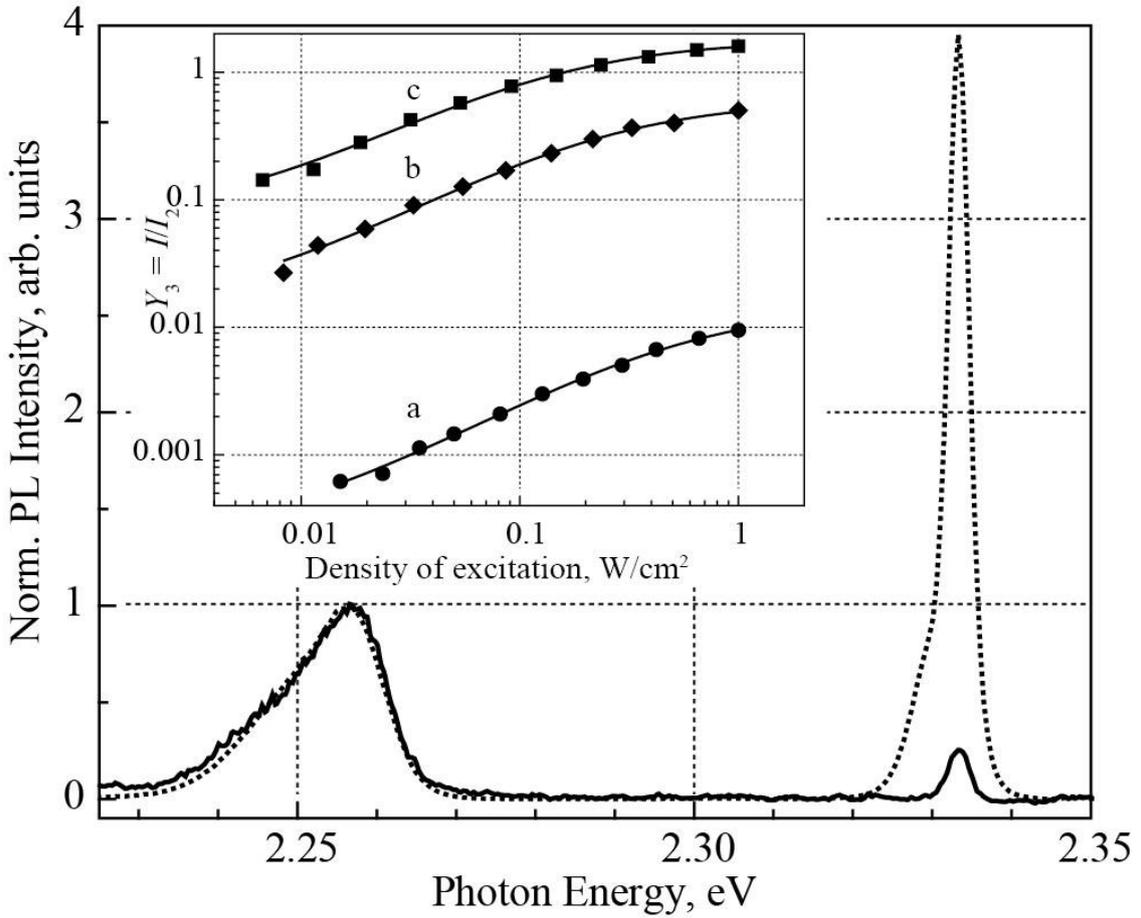

Fig.1. Normalized PL spectra of the sample with 75 ML barrier thickness excited at 2.8 eV ($T = 5$ K) for density of excitation $J = J_0 = 1$ W/cm$^2$ and $J = 0.015\ J_0$. (dashed and solid lines, respectively). The inset shows the dependences of $Y_3(J) = I_1/I_2$ for the samples with barrier thicknesses of 45, 65 and 75ML (symbols a, b and c, respectively). Solid lines at the inset indicate the fits by Eq.(1).

The PL spectra shown in Fig. 1 are normalized to the peak intensity of the $I_2$ band. With this representation, the main result of increasing of the excitation intensity is a rapid increase of the intensity of the $I_1$ band comparing to that of the $I_2$ band. As follows from the model proposed in [14], this behavior should be expected if layers $D_1$ and $D_2$ are tunnel-coupled. In more details, the dependence of the ratio of the integral intensities of the $I_1$ and $I_2$ bands $Y_3 = I_1/I_2$ on the density of excitation $J$ in the PL spectrum of a sample with a barrier thickness of 75 ML is shown in the inset of Fig. 1. Here, the $Y_3$ ratio of the intensities of the $I_1$ and $I_2$ bands of samples with barrier thicknesses of 45 and 65 ML are also shown. As can be seen from the data, the dependences of $Y_3(J)$ in all studied samples, are qualitatively the same. Moreover, the relative intensity of the $I_1$ bands in comparison with that of $I_2$ bands at the same excitation level increases significantly in a set of samples with barrier thicknesses of 45 – 65 – 75 ML, which obviously reflects a decrease of the tunnel transparency of barriers in this series of the samples.

In order to study the relative role of hot excitons and free carriers in the formation of PL spectra we investigated the dependence of the ratio $Y_3 = I_1/I_2$ of the integrated intensities of the emission bands from shallow and deep QWs on the excitation intensity $J$ by lasers with different wavelengths. As the example, Fig. 2 shows the experimental results on the $Y_3(J)$ dependence obtained at excitation by lasers with wavelengths of 517, 473, and 442 nm. In the model [14], it was assumed that the parameters of the capture of electrons, holes, and excitons into shallow and deep QWs are the same. In this case, the ratio of the PL intensities from shallow and deep QWs as a function of $J$ is determined by the equation [14]:

$$Y_3(J) = I_1/I_2 = [\alpha(J + \beta) + J] / [\gamma(1+\alpha)(J + \beta) + \beta], \quad (1)$$

where $\alpha$ is the ratio of the coefficients of capture of single carriers and excitons into the QW, $\beta$ is inversely proportional to tunneling time $\tau_t$ and determines the excitation intensity at which the probability of tunneling from shallow to deep QWs of a single charge carrier is 0.5, and the parameter $\gamma$ determine the ratio of the probabilities of capture of electronic excitations in a shallow and deep QWs..

Note, that the experimental conditions for excitation of PL by the sources with different wavelengths (such as absorption coefficient at the wavelength of excitation, size and shape of the excitation spot on the sample surface, etc.) can differ markedly. In this case, the characterization of the excitation intensity $J$ by the measured value of the sample illumination is impossible. For this reason, to compare the data on the dependences $Y_3(J)$ obtained at excitation with different wavelengths, it is more convenient to present the excitation intensity scale in relative units $J/J_{max}$, taking the maximum pump intensity as a unit. At such representation, the values of parameter $\beta$ in Eq. (1) cannot be easily compared between experiments with different wavelengths. However, one can see from Eq. (1) that multiplying both $\beta$ and $J$ by the same constant does not changes the $Y_3(J)$, since $Y_3(J)$ depends only on the ratio $J/\beta$. Thus, the scaling of different experimental dependences of $I_1/I_2$ along the $x$-axis is a reasonable way to compare them with each other, which is equivalent to multiplying $\beta$ by some constant.

As one can see from the data presented in Fig. 2, the experimental dependence $Y_3(J)$ is well approximated by Eq. (1). The parameter $\gamma$ with such a fit is close to unity, which is in favor of the model approximation used. As follows from [14], in the limit of high excitation intensities, the mechanism of population of radiating states in both QWs is the same and is determined mainly by the independent capture of electrons and holes in QWs. As shown in Fig. 2 for sufficiently high excitation intensities, this conclusion is consistent with the general character of the $Y_3(J)$ dependences obtained for different excitation wavelengths. The case of low excitation intensities is more interesting. As noted above, with sufficiently weak excitation, the waiting time for the capture of the second carrier may become longer than the time to tunnel of the first captured carrier into the deep QW. In this case the PL from the shallow QW will be determined by the recombination of hot excitons trapped in the QW. Thus, it should be expected that the relative intensity from the shallow QW, characterized by $Y_3$ ratio and depending primarily on the parameter α, should exponentially decrease with the increasing number of LO phonons emitted during energy relaxation.

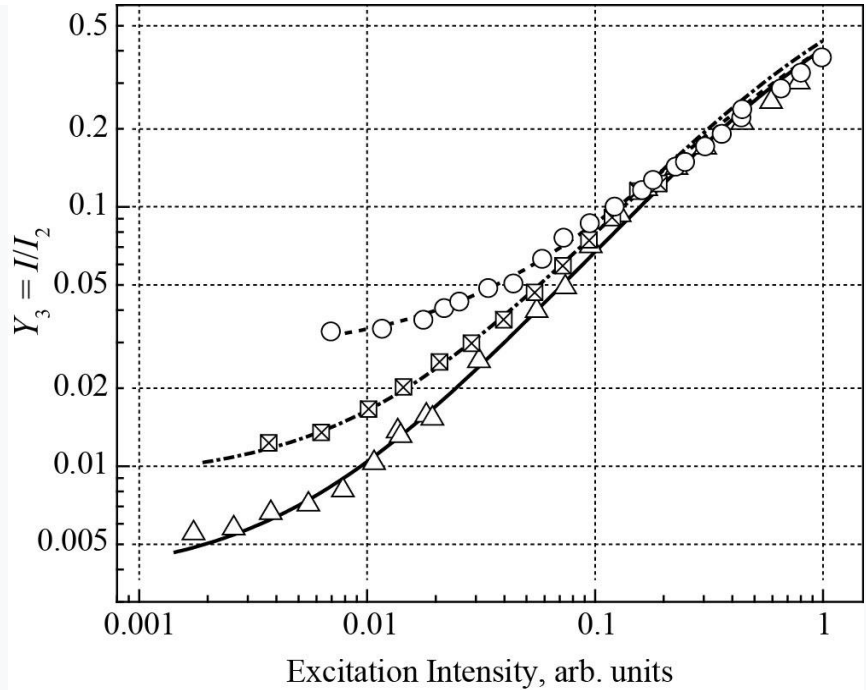

Fig. 2. Ratio of PL intensities from shallow and deep QWs versus the excitation intensity for three excitation wavelengths of 517, 473 and 442 nm (circles, crossed squares and triangles, respectively). The lines show the fits by Eq. (1) with the parameters α and γ given in the Table.

TABLE

| No | Wavelength of excitation, nm / exciting photon energy $E_0$, eV | Initial kinetic energy of e–h pairs $\Delta = (E_0 - E_g)$, eV / $n = \Delta / E_{LO}$, eV ($E_g = 2.38$ eV) | $Y_3$ ($J \sim 0$) estimated from experimental data | $\alpha$ (from Eq. (1), see text) | $\gamma$ (from Eq. (1), see text) |
|---|---|---|---|---|---|
| 1 | 441.6 / 2.807 | 0.427 / 16.5 | 0.005 | 0.0074 | 1.47 |
| 2 | 450.7 / 2.750 | 0.37 / 14.3 | 0.007±0.003 | 0.0078 | 1.0 |
| 3 | 473 / 2.621 | 0.241 / 9.3 | 0.012 | 0.018 | 1.0 |
| 4 | 507.5 / 2.443 | 0.063 / 2.4 | 0.033 | 0.08 | 1.48 |
| 5 | 517.3 / 2.396 | 0.016 / 0.6 | 0.032 | 0.0567 | 0.77 |

Indeed (Fig. 3), the value of $Y_3$ at the lowest excitation levels which allows to detect $I_1$ band is fairly well described by the dependence

$$Y_3 (J \sim 0) = a \cdot \exp(-x/b_{PL}), \qquad (2)$$

where $x = (E_0 - E_g)$ is the initial kinetic energy of hot excitons, $E_0$ is the energy of exciting photons, and $E_g$ is the band gap of ZnTe. The parameter $b_{PL}$ in Eq. (2) is equal to 0.21 eV.

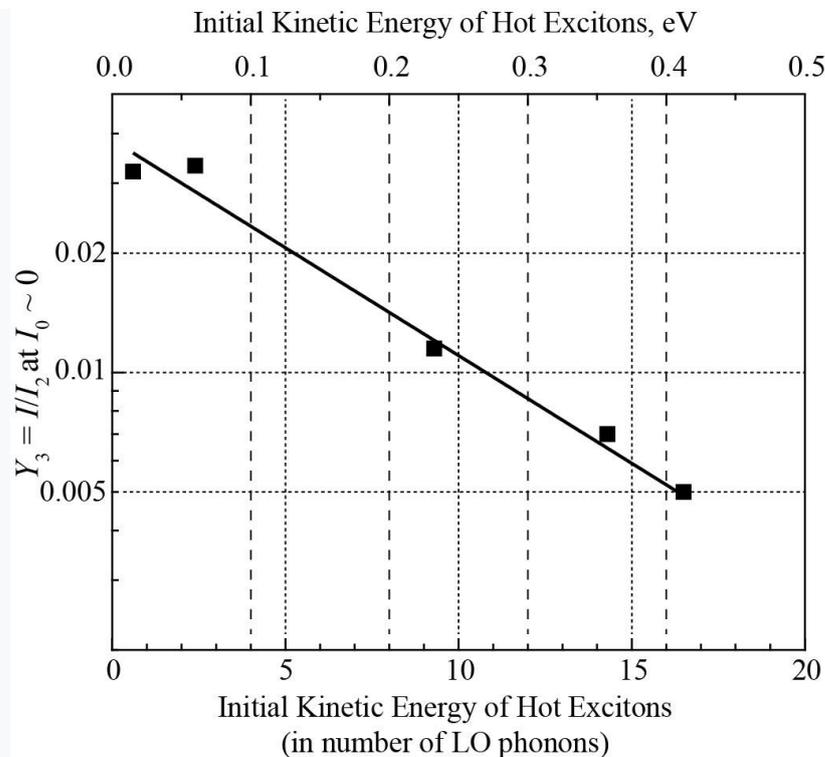

Fig. 3. Value $Y_3$ at lowest excitation intensities $J$ versus the initial kinetic energy of excitons for five laser wavelengths given in the Table. Solid line: fitting according to Eq. (2) with parameter $b_{PL} = 0.21$ eV.

Cascade relaxation of hot excitons with LO phonon emission should also be manifested in the PLE spectra. Indeed, the PLE spectra obtained with Xe lamp as an excitation source have a distinct oscillating structure. Fig. 4 shows the PL spectrum under above-barrier excitation and the PLE spectrum at detector position 2.3375 eV which corresponds to the maximum of the PL band of a shallow QW. The maxima in the PLE spectrum are spaced from each other by (26.2 ± 0.6) meV, which, with a good accuracy, coincides with the ZnTe LO phonon energy.

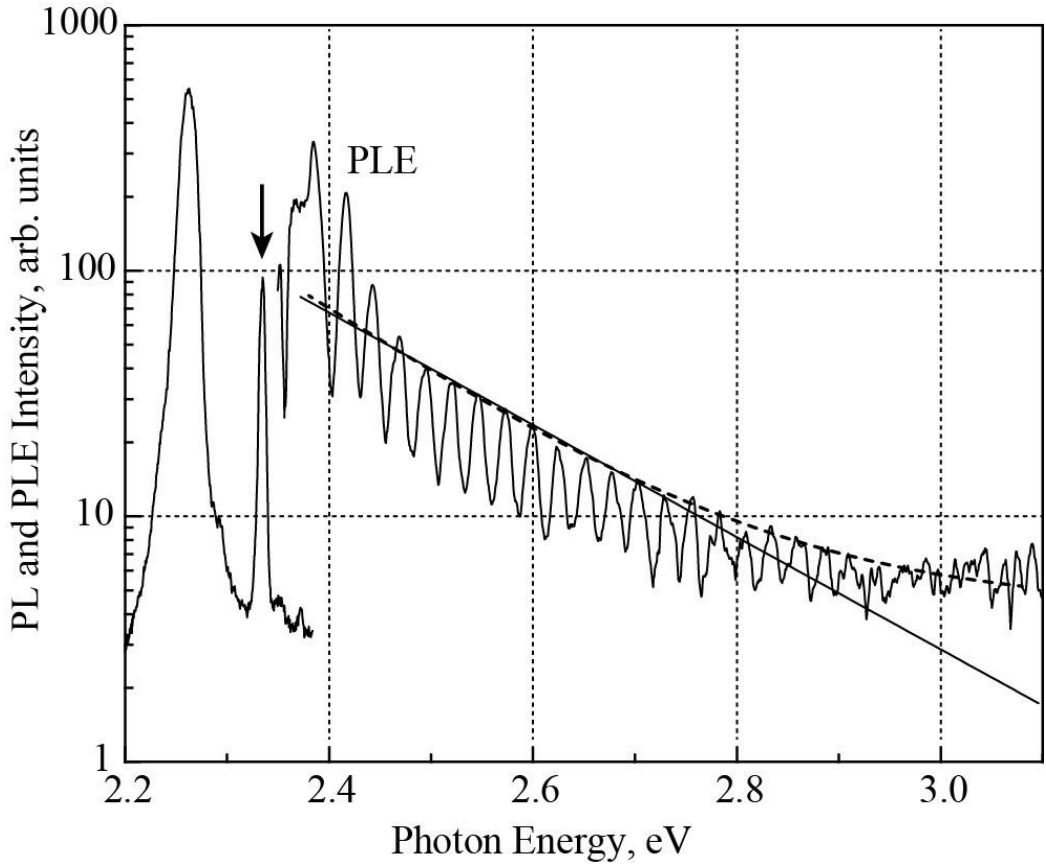

Fig. 4. PL spectrum of the sample with 65 ML ZnTe barrier thickness at excitation in the region of 3.1 eV and the PLE spectrum of the same sample at the detector position indicated by the arrow. The solid and dashed lines are fitting according to the Eqs (3) and (4) with the parameters $b_{PLE}$ = 0.19 and 0.16 eV, respectively, and $c$ = 0.057 in Eq.(4). $T$ = 4.2 K.

The intensity of the PLE spectrum decreases with increase of the energy of excitation photons. The ratio of intensity of two neighboring maxima $J_n / J_{n+1}$ practically does not depend on the numbers of $n$ and is equal to (1.15 ± 0.05). This result means that the decrease in intensity at the maxima of the PLE spectrum in the energy region $E > 2.5$ eV can be described by an exponential dependence of the form

$$J(E) = a \cdot \exp(-x/b_{PLE}) \qquad (3)$$

with the parameter $b_{PLE}$ = 0.19 eV in good agreement with the value $b_{PL}$ obtained in the dependence (2).

A decrease in the intensity of the maxima in PLE spectrum depending on the number of emitted phonons reflects decrease in the number of excitons, which, as a result of energy relaxation, are eventually captured by the states that form the PL spectrum of shallow QW. Obviously, the deviation from the exponential dependence Eq. (3) in the range $E > 2.8$ eV is determined by the contribution to the PL of the recombination of separately trapped electrons and holes. In general, the shape of the PLE spectrum in the spectral range $E > 2.5$ eV can be represented as

$$J(E) = a \left[ \exp(-x/b_{PLE}) + c \right]. \qquad (4)$$

As can be seen from Fig. 3, the deviation of decrease in the intensity of the PLE spectrum from the simple exponential dependence Eq. (3) in the region of 2.5 eV $< E <$ 2.8 eV is relatively small.

It is worth to note, that the intensities of the first few maxima in the PLE spectrum differ markedly and consistently from the Eq. (4). We assume that the excess of the PLE spectrum over Eq. (4) in the spectral range $E < 2.5$ eV is due to the dependence of the exciton creation probability on the wavelength of excitation, which, therefore, leads to the dependence of the prefactor $a$ on the energy of excitation. It was demonstrated [16] that contribution to photon absorption caused by an indirect transition with simultaneous exciton creation and phonon emission decreases rapidly with increasing excitation photon energy. On the other hand, consideration of the continuous exciton spectrum, where the electron and the hole hold their spatial correlation during energy relaxation, leads to a much smoother dependence on the excitation photon energy (see [16, 17]). As a result, the difference of the shape of PLE spectrum from the dependence defined by the Eq. (4) can be explained by the fact that for the excitation energy close to the ZnTe band gap the contribution caused by exciton creation probability through discrete exciton states dominates, while for photon energies exceeding the band gap by the energy of several LO phonons, this factor rapidly decreases and becomes negligible.

The experimental results summarized above in the Figs 2–4 were illustrated by 65 ML ZnTe barrier sample, but similar behavior was obtained also in 45 and 75 ML barrier samples.

## 4. Conclusion

We studied the luminescence spectra of a set of CdTe/ZnTe heterostructures formed by ultrathin CdTe QWs with nominal thicknesses of 2 ML and 4 ML separated by ZnTe barriers of different thicknesses. It was found that up to the barrier thickness of 75 ML the ratio of PL intensities from shallow ($I_1$) and deep ($I_2$) CdTe QWs depends on the intensity of excitation. The detected dependence of the $I_1/I_2$ ratio on the intensity of excitation can be well described by the model proposed in [14], which means that the tunnel probability between QWs is different for isolated charge carriers and excitons.

We have found that PLE spectra of a shallow QW has a pronounced oscillating structure with a period close to the energy of ZnTe LO phonon. It is concluded that this observation, as well as exponential decrease of PLE intensity with an increase in the number of emitted phonons, are characteristics of the cascade relaxation of hot excitons. The contribution of hot excitons to the PL from shallow QW is studied as a function of the initial kinetic energy of electronic excitations created by light absorption, and it is shown that this dependence also corresponds to the model of energy relaxation of hot excitons.

**Declaration of competing interest**

The authors declare that they have no known competing financial interests or personal relationships that could have appeared to influence the work reported in this paper.

**Credit authorship contribution statement**

V. Agekyan: Investigation, Writing — review & editing. G. Budkin: Conceptualization, Methodology, Writing — review & editing. M. Chukeev: Investigation. N.Filosofov: Investigation, Writing — review & editing. G.Karczewski: growth of the samples, Writing — review & editing. A. Serov: Investigation, Writing — review & editing. A. Reznitsky: Conceptualization, Investigation, Writing — original draft, Writing — review & editing.

**Acknowledgments**

This research was supported within the State Assignments from the Ministry of Science and Higher Education of the Russian Federation to the Ioffe Institute (0040-2019-0006), the St. Petersburg State University project INI 2019 id 36463378, and National Science Center (Poland) (project no. 2018/30/M/ST3/00276).